\documentclass[twocolumn,showpacs,preprintnumbers,amsmath,amssymb,epsfig,widetext]{revtex4-1}

\usepackage{graphicx}
\usepackage{dcolumn}
\usepackage{bm}
\usepackage{epsfig}
\usepackage{color}

\def\la{\mathrel{\mathpalette\fun <}}

\def\fun#1#2{\lower3.6pt\vbox{\baselineskip0pt\lineskip.9pt
  \ialign{$\mathsurround=0pt#1\hfil##\hfil$\crcr#2\crcr\sim\crcr}}}
\def\simgt{\mathrel{\lower0.6ex\hbox{$\buildrel {\textstyle >}
 \over {\scriptstyle \sim}$}}}
\def\simlt{\mathrel{\lower0.6ex\hbox{$\buildrel {\textstyle <}
 \over {\scriptstyle \sim}$}}}

\input epsf

\newcommand{\hompc}{\,h\,{\rm Mpc}^{-1}}
\newcommand{\mpcoh}{\,h^{-1}\,{\rm Mpc}}
\newcommand{\mnras}{MNRAS}
\newcommand{\apjs}{ApJS}
\newcommand{\apjl}{ApJL}
\newcommand{\aj}{AJL}

\def\be{\begin{equation}}
\def\ee{\end{equation}}
\def\ba{\begin{eqnarray}}
\def\ea{\end{eqnarray}}

\def\nn{\nonumber}

\begin{document}

\preprint{}

\title{Measuring Coherent Motions in the Universe}

\author{Yong-Seon Song$^{1,2}$\footnote{ysong@kias.re.kr}, Cristiano G. Sabiu$^3$, Issha Kayo$^{4,1}$, Robert C. Nichol$^{1}$}
\affiliation{$^1$Institute of Cosmology $\&$ Gravitation, Dennis Sciama Building,
University of Portsmouth, Portsmouth, U.K. \\
$^2$Korea Institute for Advanced Study, Dongdaemun-gu, Seoul 130-722, Korea\\
$^3$Department of Physics \& Astronomy, University College London,
Gower Street, London, U.K. \\
$^4$ Institute for the Physics and Mathematics of the Universe, University of Tokyo, 5-1-5 Kashiwanoha, Chiba 277-8583, Japan}

\date{\today}

\begin{abstract}

We present new measurements of the coherent motion of galaxies based on observations of the large-scale redshift--space distortions seen in the two--dimensional two--point correlation function of Luminous Red Galaxies in Data Release Seven of the Sloan Digital Sky Survey. We have developed a new methodology for estimating these coherent motions, which is less dependent on the details of galaxy bias and of the cosmological model to explain the late--time acceleration of the expansion of the Universe. We measure a one--dimensional velocity dispersion of galaxies on large--scales of $\sigma_v=3.01^{+0.45}_{-0.46} \mpcoh$  and $\sigma_v=3.69^{+0.47}_{-0.47} \mpcoh$ at a mean redshift of $z=0.25$ and $0.38$ respectively. These values are fully consistent with predictions for a WMAP7--normalised $\Lambda$CDM Universe and inconsistent at confidence of $3.8\sigma$ with a Dvali-Gabadadze-Porrati (DGP) model for the Universe. We can convert the units of these $\sigma_v$ measurements to  $270^{+40}_{-41}$ km/s and $320^{+41}_{-41}$ km/s respectively (assuming a $\Lambda$CDM universe), which are lower that expected based on recent low redshift ($z<0.2$) measurements of the peculiar velocity field (or ``bulk flows"). It is difficult to directly compare these measurements as they cover different redshift ranges and different areas of the sky. However, one possible cosmological explanation for this discrepancy is that our Galaxy is located in unusually over, or under, dense region of the Universe.



\end{abstract}

\pacs{draft}


\maketitle

\section{Introduction}

The last decade has seen a revolution in cosmology with the emergence of a standard model for the Universe dominated by a strange substance known as ``dark energy", with an effective negative pressure. Since the first evidence of dark energy in 1998 ~\cite{Perlmutter:1998np,Riess:1998cb}, there has been substantial observational and theoretical research aimed at understanding the true nature of this phenomenon. In recent years, many authors have started exploring the possibility that dark energy, and the observed acceleration of the expansion of the Universe, could be the consequence of an incomplete theory of gravity on cosmological scales and may require modifications to Einstein's  theory of General Relativity. 

One of the most direct methods of testing our assumed theory of gravity is to perform consistency checks between the geometrical expansion history of the Universe, as measured by cosmological probes like Type Ia Supernovae and the Baryon Acoustic Oscillations, and the evolution of density inhomogeneities in the Universe~\cite{Song:2005gm,ishak05,knox05,linder05,Jain:2007yk}.  The growth of structures in the Universe can be measured using a variety of techniques, but a popular and direct method involves measuring the coherent peculiar velocities of galaxies on large scales (i.e., the motion of galaxies after the cosmological expansion has been removed) caused by their infall into large scale overdensities like clusters and superclusters of galaxies~\cite{Song:2008vm,Song:2009zz}. Traditionally, the local peculiar velocity field, or ``bulk flow'' of galaxies, has been estimated using samples of galaxies where the peculiar velocity of each galaxy has been determined using secondary distance indicators (e.g. the Tully--Fisher relationship \cite{1998AJ....116.2632G,1999ApJ...522..647W,2007ApJS..172..599S} or Fundamental Plane \cite{2000ApJ...537L..81D,2002AJ....123.2990B,2003AJ....126.2268W}). Alternatively, the coherent motion, or flow, of galaxies can be statistically estimated from their effect on the clustering measurements of large redshift surveys, or through the measurement of redshift--space distortions.

Historically, measurements of the local galaxy peculiar velocity field have been plagued by systematic uncertainties \cite{2007MNRAS.375..691S}, and small sample sizes, which have made their cosmological constraints uncompetitive compared to other probes of the Universe. However, a recent re--analysis of these earlier peculiar velocity galaxy surveys~\cite{Watkins:2008hf,Feldman:2009es,Lavaux:2008th} has now provided a consistent observational picture from these data, and finds significant evidence for a larger than expected bulk flow in the local Universe (within 100 $\mpcoh$). This new analysis is also consistent with new measurements of the local bulk flow using the kinetic Sunyaev-Zeldovich effect \cite{1972CoASP...4..173S} of massive clusters of galaxies out to at least 800 Mpc~\cite{Kashlinsky:2008ut,Kashlinsky:2009dw}, which leads to the intriguing situation that these local measurements appear to be significantly greater in amplitude, and on larger scales, than expected in a concordance, WMAP7--normalised $\Lambda$CDM cosmological model. Such discrepancies with $\Lambda$CDM may give support to theories of modified gravity~\cite{Afshordi:2008rd}.

Given the importance of these local large--scale bulk flow measurements, we investigated in a previous paper \cite{2010JCAP...01..025S} the likelihood of such large coherent flows using an alternative methodology based upon measurements of the redshift--space distortions seen in the clustering of a sample of galaxy clusters selected from the Sloan Digital Sky Survey (SDSS)~\cite{York:2000gk}. We detected no statistical evidence for the coherent flows of the same size, and scale, as discussed above, but due to the sample size the statistical errors were large (see~\cite{2010JCAP...01..025S} for details). Therefore, in this paper, we re-visit our earlier work using a larger cosmological data--set and recent theoretical improvements in the modelling of redshift--space distortions. The goal of this paper is to statistically study the likelihood of the local bulk flow measurements within a large cosmological volume of the Universe, well beyond these nearby measurements.

In section II of this paper, we outline in greater detail the fitting techniques used in \cite{2010JCAP...01..025S} to formulate the line--of--sight smearing effect due to the coherent bulk flow of galaxies. In section III, we explain our data analysis of a sample of SDSS Luminous Red Galaxies (LRGs; \cite{2001AJ....122.2267E}) and our measurement of their two--dimensional redshift--space correlation function. In section IV, we present our results from comparing our LRG measurements to our model for redshift--space distortions, and compare our results to cosmological predictions and previous observations. We conclude in section V.

\section{Modelling Redshift--Space Distortions}
The aim of this work is to statistically measure the coherent motion of galaxies on large scales. That is, the linear velocity of galaxies excluding both the cosmological Hubble expansion and any smaller scale non--linear components.

The coherent motions of galaxies introduce redshift--space distortions -- an anisotropic feature -- into the measured clustering statistics.
As originally proposed in ~\cite{Kaiser:1987qv}, a distant observer should expect a multiplicative enhancement of the overdensity field along the line--of--sight, compared to the transverse direction, due to such coherent peculiar motion of galaxies. This ``Kaiser effect", as it is now know, can be seen as a ``squashing" or flattening of the two--dimensional two--point correlation function ($\xi_s(\sigma,\pi)$), where the correlation function is decomposed into two vectors; one parallel to the line--of--sight ($\pi$) and the other perpendicular to the line--of--sight ($\sigma$). Information about the coherent velocities of galaxies can then be extracted from the two--dimensional correlation function via careful theoretical modelling of these redshift--space distortion effects. 

In this paper, we utilise several theoretical improvements to the original redshift--space distortion work given in~\cite{Kaiser:1987qv}. Throughout this paper, we refer to the original work of~\cite{Kaiser:1987qv} as the ``Kaiser limit'', which formulated the coherent motions of galaxies as an additional Dopper shift to the cosmological redshift. The ``Kaiser limit" is suitable to test theoretical models in the linear region of galaxy perturbations (on large scales), but over the last decade there have been significant improvements to the modelling of these effects. Below, we repeat some of the formalism presented in our first paper \cite{2010JCAP...01..025S} to aid the reader and provide continuity with additional work presented in following subsections. 

\subsection{Representation of the correlation function in the Kaiser limit}

The observed power spectrum of density fluctuations in redshift--space, $\tilde{P}_{\rm ob}(k,\mu)$, can be written as,
\ba\label{eq:ps}
\tilde{P}_{\rm ob}(k,\mu)&=&\left[P_{\delta_g\delta_g}(k) + 2\mu^2P_{\delta_g\Theta}(k) + \mu^4P_{\Theta\Theta}(k)\right] \nn  \\
&\times&G(k,\mu;\sigma_v)\,,
\ea
where $\delta_g$ and $\Theta$ denote galaxy density and velocity fields respectively (where $\Theta=\theta/aH$ and $\theta$ is the divergence of the velocity fields). In Eq.~\ref{eq:ps}, the function $G$ denotes the additional suppression effect due to coherent motions which affects the galaxy--galaxy, velocity--velocity and galaxy--velocity power spectra equally, and is a function of separation ($k$), angle between the galaxies (denoted by $\mu$) and has a given smoothing scale, or velocity dispersion ($\sigma_v$), which is the parameter at the heart of our analysis and this paper. 

As discussed in our previous paper, we can decompose these power spectra into a scale--dependent ($D$) and scale--independent part ($g$) given by,
\ba\label{eq:pall}
P_{\Phi\Phi}(k,a)&=&D_{\Phi}(k)g_{\Phi}^2(a),\nn\\
P_{\delta_g\delta_g}(k,a)&=&D_{m}(k)g_{b}^2(a),\nn\\
P_{\Theta\Theta}(k,a)&=&D_m(k)g_{\Theta}^2(a),
\ea
where the subscript $\Phi$ denotes the curvature perturbation in the Newtonian gauge, 
\begin{equation}
 ds^2=-(1+2\Psi)dt^2+a^2(1+2\Phi)dx^2\,.
\end{equation}
The growth function $g_b$ is defined as $g_{b}\equiv b\,g_{\delta_m}$ where $b$ is the standard linear bias parameter between the density of galaxies and the underlying dark matter, $\delta_m$. As we follow the positive sign conversion, $g_{\Theta}$ is the growth function of $-\Theta$.

The shape factor of the perturbed metric power spectra $D_{\Phi}(k)$ is defined as
\be
D_{\Phi}(k)=\frac{2\pi^2}{k^3}\frac{9}{25}\Delta^2_{\zeta_0}(k)T^2_{\Phi}(k),
\ee
which is a dimensionless metric power spectra at $a_{eq}$ (the matter--radiation equilibrium epoch), and $\Delta^2_{\zeta_0}(k)$ is the initial fluctuations in the comoving gauge and  $T_{\Phi}(k)$ is the transfer function normalised at $T_{\Phi}(k\rightarrow 0)=1$. The primordial shape $\Delta^2_{\zeta_0}(k)$ depends on $n_S$ (the slope of the primordial power spectrum), as $\Delta^2_{\zeta_0}(k)=A^2_S(k/k_p)^{n_S-1}$, where $A^2_S$ is the amplitude of the initial comoving fluctuations at the pivot scale, $k_p=0.002$ ${\rm Mpc}^{-1}$. The intermediate shape factor $T_{\Phi}(k)$ depends on $\omega_m$ ($\omega_m\equiv\Omega_mh^2$). 

The shape factor for the matter fluctuations, $D_m(k)$, which is important for both the galaxy--galaxy and velocity--velocity power spectra in Eq. \ref{eq:pall} above, is given by the conversion from $D_{\Phi}(k)$ of,
\be
D_m(k)\equiv\frac{4}{9}\frac{k^4}{H_0^4\Omega_m^2}D_{\Phi}(k),
\ee
where, assuming $c=1$, we can write $H_0\equiv 1/2997\hompc$. 

In the Kaiser limit, we would thus assume $P(k)\equiv P^{\rm lin}(k)$, where $P^{\rm lin}(k)$ is the linear power spectrum, and no smearing of the large scale power spectra due to the random motions of galaxies within individual dark matter halos, or the ``Finger--of--God'' (FoG) effect seen on small--scales in redshift surveys. Therefore, the observed compression of $\xi_s(\sigma,\pi)$ along the line--of--sight due to coherent infall of galaxies  around large--scale structures can be written in configuration space as,
\ba\label{eq:xis}
\xi_s(\sigma,\pi)(a)&=&
\left(g_b^{2}+\frac{2}{3}g_bg_{\Theta}+\frac{1}{5}g_{\Theta}^{2}\right)\xi_0(r){\cal
  P}_0(\mu)\nn\\
&-&\left(\frac{4}{3}g_bg_{\Theta}+\frac{4}{7}g_{\Theta}^{2}\right)\xi_2(r){\cal
  P}_2(\mu)\nn\\
&+&\frac{8}{35}g_{\Theta}^{2}\xi_4(r){\cal P}_4(\mu),
\ea
where ${\cal P}_l(\mu)$ is the Legendre polynomial and the spherical
harmonic moment $\xi_l(r)$ is given by,
\be
\xi_l(r)=\int\frac{k^2dk}{2\pi^2}D_m(k)j_l(kr),
\ee
where $j_l$ is a spherical Bessel function. 

\begin{figure*}[t]
\centering
\includegraphics[width=0.47\textwidth]{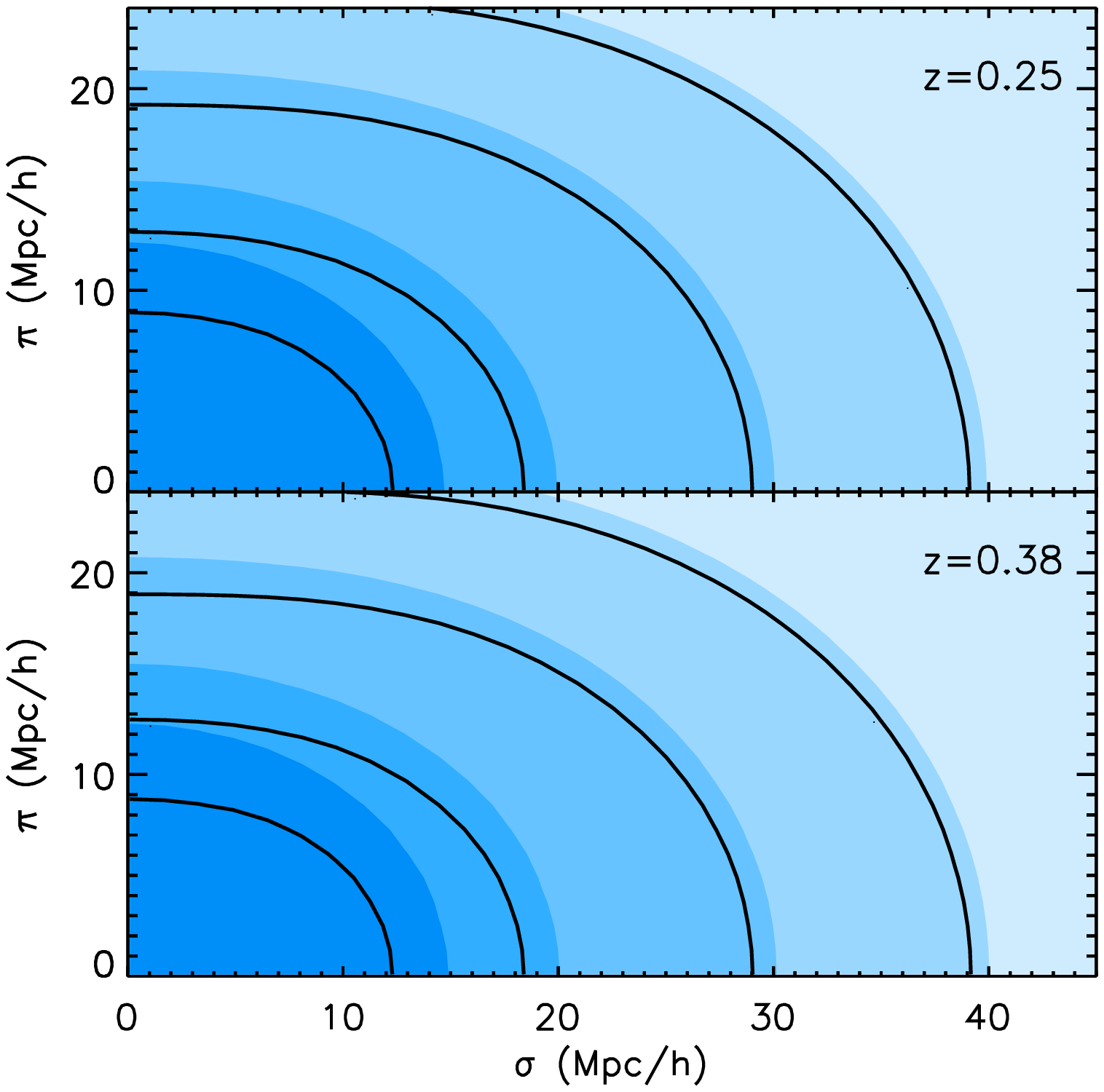}\hfill
\includegraphics[width=0.47\textwidth]{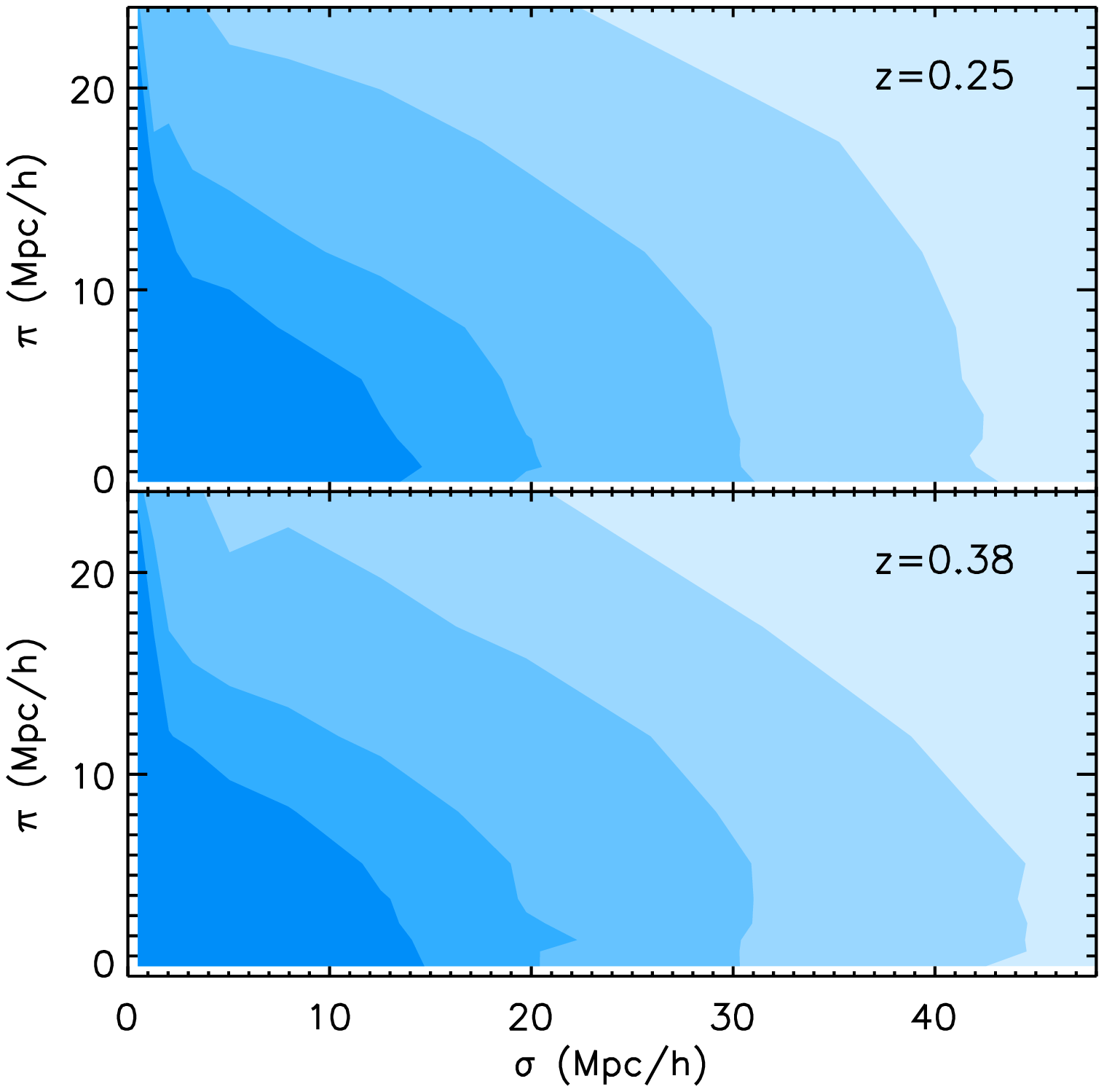}
\caption{\footnotesize {\it (Left panel)} The theoretical 2D--correlation function $\xi_s(\sigma,\pi)$ in the Kaiser limit (unfilled black contours) and using our modified Kaiser effect (filled blue contours). Levels of the contours are $\xi_s(\sigma,\pi)=(1,0.5,0.2,0.1)$ from the inner to outer contour. Contours in the upper panel are derived from $\Lambda$CDM model with density bias $b=1.82$ at $z=0.25$, and contours in the lower panel are derived from $\Lambda$CDM model with density bias $b=1.92$ at $z=0.38$. {\it (Right panel)} The observed 2D--correlation function $\xi_s(\sigma,\pi)$ from SDSS DR7 LRG catalogue with the same contour levels.}
\label{fig:sigma_pi}
\end{figure*}

\subsection{CMB priors}

The coherent evolution parts in Eq.~\ref{eq:pall} are not generally parameterised by known standard cosmological parameters. We thus normalise these growth factors at $a_{eq}$ such that,
\ba
g_{\Phi}(a_{eq})&=&1,\nn\\
g_{\delta_m}(a_{eq})&=&a_{eq}g_{\Phi}(a_{eq}), \nn\\
g_{\Theta}(a_{eq})&=&\frac{dg_{\delta_m}(a_{eq})}{d\ln a}\,.
\ea
Then, we treat $g_b$ and $g_{\Theta}$ as free parameters to be measured by redshift--space distortions with  CMB priors used to determine the shape factor $D_m(k)$ at the last scattering surface.

The shape factor $D_m(k)$ depends mainly on the following set of cosmological parameters, $\omega_m,\,n_S,\,A_S$ and $\Omega_m$. While the first two parameters ($\omega_m$ and $n_S$) are well constrained by the CMB at the last scattering surface (regardless of theoretical model used to describe the late--time acceleration of the expansion of the Universe), the last two parameters ($A_S$ and $\Omega_m$) can not be determined by the CMB alone. The primordial amplitude $A_S$ has been measured to an accuracy of less than $5\%$ from WMAP7 alone, however this constraint becomes weaker due to the unknown reionization history of the Universe (tighter constraints are achieved by assuming a stepwise parameterisation of the reionization history). For the matter content of the Universe, the CMB measures $\omega_m$, not $\Omega_m$, which depends on the Hubble Constant.

Therefore, given our CMB priors, $D_m(k)$ can be expressed as,
\be
D_m(k) = \left(\frac{A_S}{\Omega_m}\frac{\Omega_m^*}{A_S^*}\right)^2 D^*_m(k:A_S^*,\Omega_m^*)\,,
\ee
where $A_S^*$ and $\Omega_m^*$ are specific reference values. We use the best fit cosmological parameters of the WMAP7--normalised $\Lambda$CDM model in this paper, $A_S^*=\sqrt{2.41\times 10^{-9}}$ and $\Omega_m^*=0.264$. Then the power spectra can be  re--written as,
\ba
P_{\delta_g\delta_g}(k,a)&=&\left(g_b\frac{A_S}{\Omega_m}\frac{\Omega_m^*}{A_S^*}\right)^2 D^*_m(k:A_S^*,\Omega_m^*),\nn\\
P_{\Theta\Theta}(k,a)&=&\left(g_{\Theta}\frac{A_S}{\Omega_m}\frac{\Omega_m^*}{A_S^*}\right)^2 D^*_m(k:A_S^*,\Omega_m^*)\,.
\ea
Here, we define new growth functions as,
\ba
g_{b}^*(a)&\equiv&g_{b}(a)\frac{A_S}{\Omega_m}\frac{\Omega_m^*}{A_S^*}\nn\\
g_{\Theta}^*(a)&\equiv&g_{\Theta}(a)\frac{A_S}{\Omega_m}\frac{\Omega_m^*}{A_S^*}\,.
\ea
Then the power spectra become
\ba
P_{\delta_g\delta_g}(k,a)&=&g_b^{*\,2} D^*_m(k:A_S^*,\Omega_m^*),\nn\\
P_{\Theta\Theta}(k,a)&=&g_{\Theta}^{*\,2} D^*_m(k:A_S^*,\Omega_m^*),
\ea
in which the power spectra are split into two parts, i.e., $g_b^*$ and $g_{\Theta}^*$ are determined through our redshift--space distortion measurements, and $D_m^*$ is given by the CMB priors and the reference values of $A_S^*$ and $\Omega_m^*$. 

Thus, we re-write the correlation function in Eq.~\ref{eq:xis} in terms of our new fitting parameters $g_b^*$ and $g_{\Theta}^*$,
\ba\label{eq:xis*}
\xi_s(\sigma,\pi)(a)&=&
\left(g_b^{*\,2}+\frac{2}{3}g_b^*g_{\Theta}^*+\frac{1}{5}g_{\Theta}^{*\,2}\right)\xi_0^*(r){\cal
  P}_0(\mu)\nn\\
&-&\left(\frac{4}{3}g_b^*g_{\Theta}^*+\frac{4}{7}g_{\Theta}^{*\,2}\right)\xi_2^*(r){\cal
  P}_2(\mu)\nn\\
&+&\frac{8}{35}g_{\Theta}^{*\,2}\xi_4^*(r){\cal P}_4(\mu),
\ea
where the spherical harmonic moment $\xi_l^*(r)$ is given by,
\be\label{eq:jl}
\xi_l^*(r)=\int\frac{k^2dk}{2\pi^2}D_m^*(k:A_S^*,\Omega_m^*)j_l(kr).
\ee
We measure $g_{\Theta}^*$ with a normalised $D^*_m(k:A_S^*,\Omega_m^*)$, not $g_{\Theta}$ paired with $D_m(k)$. However, as $P_{\Theta\Theta}=g_{\Theta}^{2} D_m(k)=g_{\Theta}^{*\,2} D^*_m(k:A_S^*,\Omega_m^*)$, $P_{\Theta\Theta}$ does not depend on any normalisation, i.e., $P_{\Theta\Theta}$ is invariant under the transformation between the $^*$ (asterisk) variables and those without above. Thus, we present measured coherent motions not in terms of $g_{\Theta}^*$, but in terms of velocity dispersion derived from $P_{\Theta\Theta}$. Hereafter, we drop {\it $`*'$} symbol, i.e. {\it `unasteriked'} quantities imply {\it `asteriked'} quantities from now.

\subsection{Modifying the Kaiser effect}

It has been pointed out in \cite{Scoccimarro:2004tg} that the description of $\xi_s(\sigma,\pi)$ on large scales (in linear theory) will be modified due to dispersion effects in the $\pi$ direction. This effect was recently investigated using $N$-body simulations and was found to be important to include~\cite{Taruya:2009ir}. Therefore, in this paper, we have modified the Kaiser limit using a Gaussian velocity dispersion term for $G$ in Eq. 1, which we will refer to as the ``modified Kaiser effect" throughout. This modified model has been shown to work reasonable well on  mock catalogues of a $\Lambda$CDM Universe (see~\cite{Taruya:2009ir,Song:2010bk,Jennings:2010uv}). 

In detail, we can write $G$ as
\ba\label{eq:vdispersion}
G(k,\mu,\sigma_v)=e^{-k^2\mu^2\sigma_v^2},
\ea
where $\sigma_v$ is the 1--D velocity dispersion on large scales, and is the parameter of interest in this paper. We can write $\sigma_v$ as,
\ba
\sigma_v^2=\frac{1}{6\pi^2}\int
P^{\rm lin}_{\Theta\Theta}(k,z) dk, \label{eq:sigma_v}
\ea
where we see that $\sigma_v$ has units of $\hompc$ as $P^{\rm lin}_{\Theta\Theta}$ and $k$ have units of $(\mpcoh)^3$ and $\hompc$ respectively. This is somewhat confusing given $\sigma_v$ is called a ``velocity dispersion"  in previous  literature, but we retain this terminology here to remain consistent with this literature. However, we note that we can convert the units of $\sigma_v$ to km/s using $aH(a)\sigma_v$, but this requires a precise measurement of the expansion history of the Universe (at the redshifts of interest), in addition to the determined $P^{\rm lin}_{\Theta\Theta}$ from our method. 

We therefore modify the Kaiser effect using $G$ defined in Eq.~\ref{eq:vdispersion}, but keeping $P(k)=P^{\rm lin}(k)$. Then the integration of $\xi_l$ in Eq.~\ref{eq:xis} should be re--expressed, as $G$ changes with varying $g^*_{\Theta}$ altering $\sigma_v$ in Eq.~\ref{eq:sigma_v} to,
\ba\label{eq:modxis}
\xi_l^*(r)=\int\frac{k^2dkd\mu_k}{(2\pi)^2} D_m^*(k)e^{-(k\mu_k\sigma_v)^2}\cos{(kr\mu_k)}{\cal P}_l(\mu_k)\,.
\ea
Here $\mu_k$ is the cosine of angle between $\vec{k}$ and the pairwise orientation.


For illustrative purposes, in the left--hand panel of Fig.~\ref{fig:sigma_pi}, we present our modelling of $\xi_s(\sigma,\pi)$ using the standard Kaiser limit (unfilled black contours) compared to the modified Kaiser effect (filled blue contours) discussed above. In these examples, we have used a galaxy bias of $b=1.86$ and $1.88$ at $z=0.25$ (upper panel) and 0.38 (lower panel) respectively (These values of $b$ were chosen to correspond to our estimated bias values from Table I to illustrate our point).  The additional suppression in the modified Kaiser effect can be clearly seen on large--scales (in the linear regime) at approximately the 10 to 20$\%$ level.

\section{Data Analysis}

\begin{figure}[t]
\begin{center}
\includegraphics[width=0.47\textwidth]{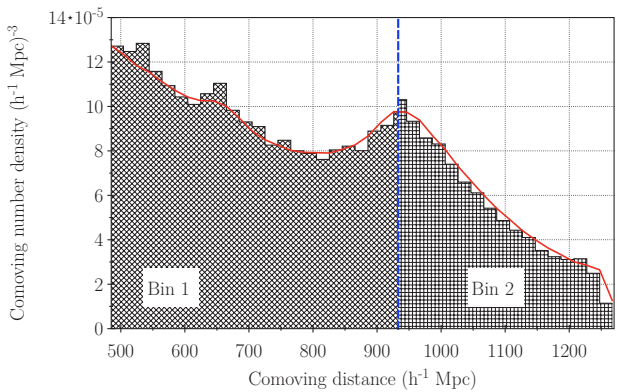}
\caption{\label{fig:zdist}\footnotesize The normalized distribution of 105,831 SDSS LRG galaxies over--plotted with the random catalogue as a smooth red line (1,664,948 points). The LRGs are split into two redshift bins at the median redshift of the whole distribution ($z=0.32$) as shown by the blue, dashed line. The low redshift bin spans the range $0.16<z<0.32$, while the higher redshift bin covers $0.32<z<0.47$. }
\end{center}
\end{figure}

\subsection{SDSS LRG sample}

In this paper, we have analysed LRGs taken from Data Release Seven (DR7) of the SDSS~\cite{2009ApJS..182..543A}. In particular, our LRG sample contains approximately 100,000 galaxies in the redshift range $0.16 < z < 0.47$ with magnitudes $-23.2 < M_g < -21.2$ 
The 3-dimensional size of the survey is best described by the effective volume~\cite{Fel94}
\begin{equation}\label{eq:Veff}
V_{eff}(k) = \int d^3r\;\left(n(\vec{r})P(k)\over 1+n(\vec{r})P(k)\right)^2
\end{equation}
where $n(\vec{r})$ is the comoving number density at every point $\vec{r}$ and $P(k)$ is the power amplitude at wavelength $k$.
The SDSS DR7 LRG sample used in this analysis has $V_{\small eff}\approx 1 (h^{-1}~{\rm Gpc})^3$. The LRGs are more biased than main sample galaxies and thus have a higher clustering amplitude which boosts the signal-to-noise on their measurements. It is this large volume, and the fact that LRGs are highly biased tracers of the underlying mass distribution, which makes them an excellent population for studying the statistics of large scale structure.

In detail, we use the publicly available LRG data 
from ~\cite{2005AJ....129.2562B, 2008ApJS..175..297A, 2008ApJ...674.1217P, 2010ApJ...710.1444K}, that include random catalogues and galaxy weights that are essential for accommodating the angular mask in such large scale galaxy surveys. We have converted the observed coordinates of LRGs (RA, Dec, redshift) into comoving cartesian distance coordinates ($D_x$, $D_y$, $D_z$) using the WMAP7 mean cosmological parameters 
based upon a $\Lambda$CDM model. Therefore, $\xi_{l,n}(r)$ in Eq.~\ref{eq:modxis} can be re--expressed using a mapping from the fiducial WMAP7 normalised $\Lambda$CDM model of $k_{\rm fid}=kD^V/D^V_{\rm fid}$ (where $D^V=(D_A^2H^{-1})^{1/3}$).

We note that by assuming a cosmological model for the conversion of redshifts and angles to distances could introduce an additional anisotropic signal in the correlation function i.e. the ``Alcock--Paczynski'' effect~\cite{Alcock:1979mp}. The size of this effect is smaller than the Kaiser effect on the scales of interest herein, e.g., if we vary our background cosmology, within reasonable limits allowed by the WMAP analysis, we see no significant difference in our correlation functions. However, we note that a more self--consistent approach will be required in the future (see~\cite{Matsubara:1996nf,Simpson:2009zj,Samushia:2010ki}).

In Fig. \ref{fig:zdist}, we show the co-moving number density of LRGs in our whole sample (compared to the random catalogue) and  we split this sample into two redshift bins as indicated (Bin 1: $0.16 < z < 0.32$ and Bin 2: $0.32 < z < 0.47$). We do not split the sample further as we wish to keep the thickness of the redshift shells ($\delta z\sim0.15$) larger than the scales of interest in our analysis, thus preventing any aliasing in the redshift direction.

We estimate the correlation function using the ``Landy-Szalay" estimator~\cite{1993ApJ...412...64L},
\begin{equation}
\xi(\sigma,\pi)=\frac{DD-2DR+RR}{RR},
\end{equation}
where $DD$ is the number of galaxy--galaxy pairs, $DR$ the number of galaxy-random pairs, and $RR$ is the number of random--random pairs, all separated by a distance
$\sigma\pm\Delta\sigma$ and $\pi\pm\Delta\pi$. All pairs are weighted using the minimum variance weighting scheme of \cite{1994ApJ...426...23F}. Each galaxy is assigned a weight according to,
\begin{equation}
w_i=\frac{1}{1+n_i(z)P_w},
\end{equation}
where $n_i(z)$ is the comoving space density at redshift $z$ and $P_w = 4\cdot10^{4}h^{-3}{\rm Mpc^3}$, as in \cite{2005ApJ...633..560E}. The pairs are also corrected for fiber collisions, where two or more galaxies separated by less than 55 arcseconds cannot be observed simultaneously. This is achieved using a secondary galaxy weight, proportional to the number of unobserved galaxies within the fiber radius. The pair counting was achieved using the parallel kd-tree code, {\texttt {NTROPY}} \cite{2007arXiv0709.1967G}.

In the right--hand panel of Fig.~\ref{fig:sigma_pi}, we present our measured $\xi_s(\sigma,\pi)$. The perpendicular component, $\sigma$, is binned logarithmically into ten bins from $1 \mpcoh$ to $60\mpcoh$, and the parallel component, $\pi$, is also binned logarithmically into ten bins, but from $1 \mpcoh$ to $30\mpcoh$.

We use the jack--knife method~\cite{2002ApJ...579...48S} to estimate statistical errors on $\xi_s(\sigma,\pi)$, which involved dividing the survey into $N$ sub-sections with approximate equal area (and thus volume) and then computing the mean and variance of $\xi_s(\sigma,\pi)$ from these $N$ measurements of the correlation function with the  $i^{th}$ region removed each time (where $i=1...N$). In our analysis, we use $N=100$ jack--knifed samples 
which were created using an adaption of the {\texttt{HEALPix}}, equal--area, pixelisation code
\cite{2005ApJ...622..759G}.

\begin{figure}[t]
 \begin{center}
 \epsfysize=3.2truein
 \epsfxsize=3.2truein
 \epsffile{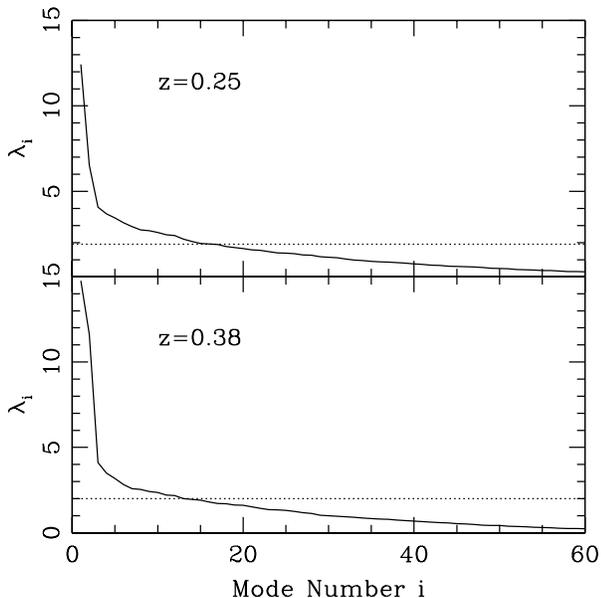}
   \caption{\label{fig:eigen}\footnotesize Eigenvalues ($\lambda_i$) at both redshift $z=0.25$ (upper panel) and $z=0.38$ (lower panel). Dotted lines represent our threshold limits.}
\end{center}
\end{figure}

The jack--knife measurements provide an estimate of the covariance matrix for all bins in our correlation function. However, this matrix can be noisy, given the number of jack--knife samples used, although previous work \cite{2009MNRAS.393.1183C, 2010arXiv1004.2244K} has shown that the number of jack--knife samples we use should provide a stable estimate of the covariance matrix. In order to reduce the statistical noise in our covariance matrix further, we follow the same procedure as described in~\cite{2010JCAP...01..025S} and ``clean'' our covariance matrix before inversion~\cite{Dekel:2001pn,Watkins:2001xe,Feldman:2003bq}. In Figure~\ref{fig:eigen}, we show the distribution of eigenvalues for the decomposition of our two covariance matrices and truncate them when the eigenvalues decrease below $\lambda_i \la 2$, which avoids any singularity in the inversion. This threshold in  $\lambda_i$ was empirically determined and, reassuringly, gives reduced $\chi^2$ values close to unity (see below). 

\subsection{Marginalisation over cosmological parameters}

\begin{figure*}[t]
\centering
\includegraphics[width=0.47\textwidth]{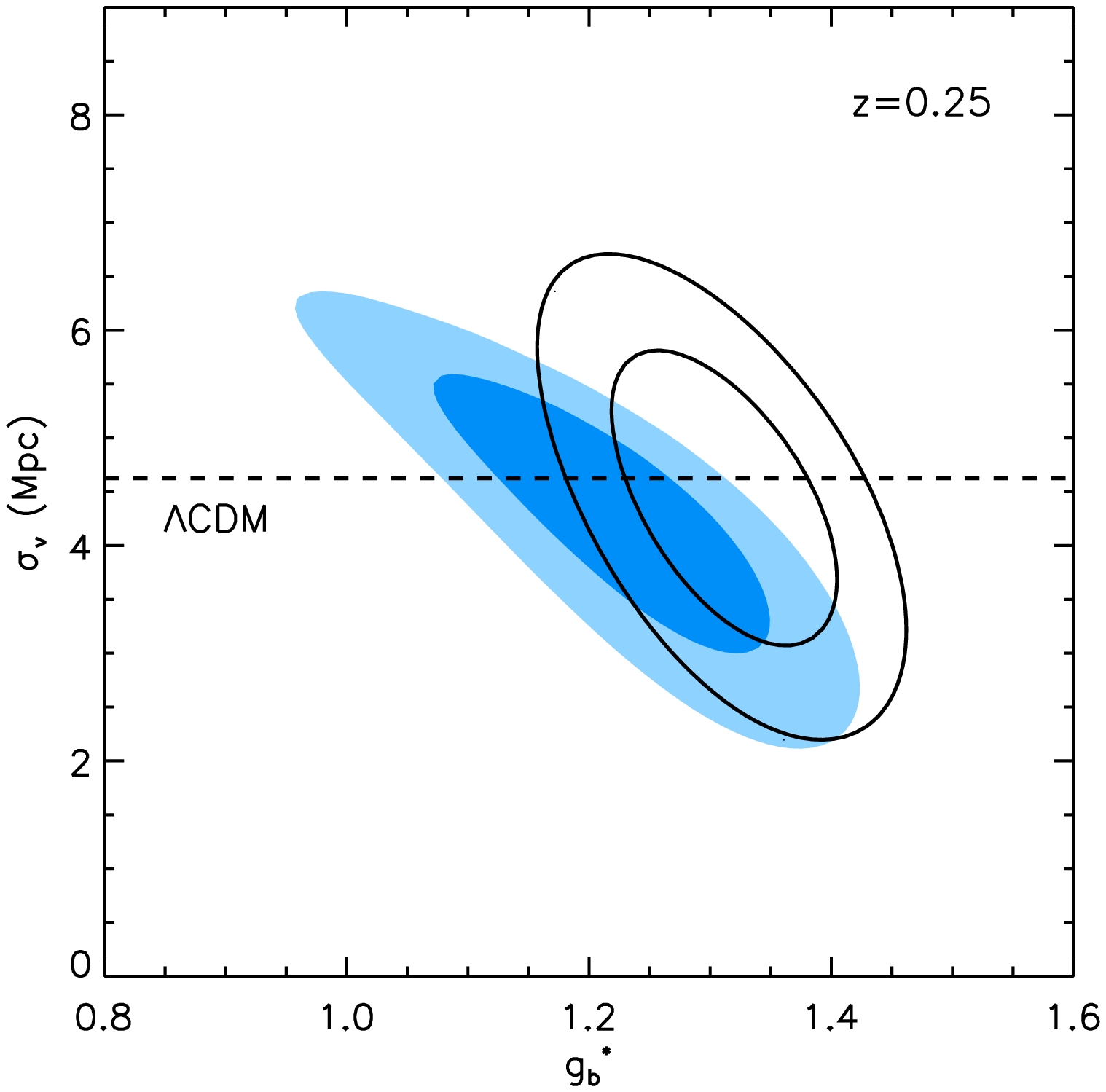}\hfill
\includegraphics[width=0.47\textwidth]{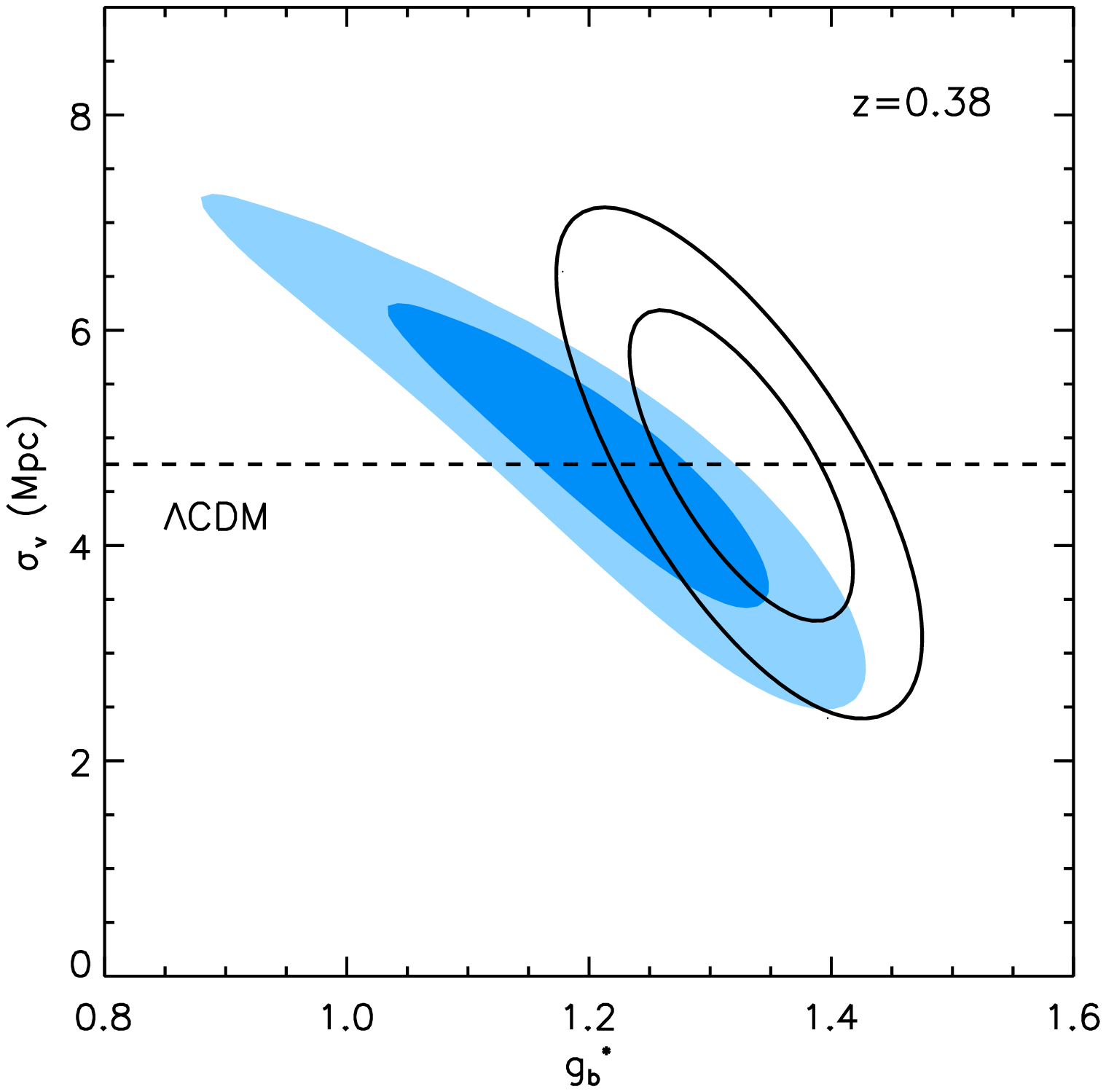}
\caption{Constraints on $g_b$ and $\sigma_v$. Unfilled black contours denote the result with Kaiser limit, while the filled blue contours are the results including the velocity dispersion effect discussed in the text. Contours are 68\% and 95\% errors.}
\label{fig:contour}
\end{figure*}

The shape of the power spectra of density fluctuations is determined before the epoch of matter--radiation equality. Under the paradigm of inflationary theory, initial fluctuations are stretched outside the horizon at different epochs which generates the tilt in the power spectrum. The predicted initial tilting is parameterised as a spectral index ($n_S$) which is just the shape dependence due to the initial conditions. When the initial fluctuations reach the coherent evolution epoch after matter-radiation equality, they experience a scale-dependent shift from the moment they re-enter the horizon to the equality epoch. Gravitational instability is governed by the interplay between radiative pressure resistance and gravitational infall. The different duration of modes during this period result in a secondary shape dependence on the power spectrum. This shape dependence is determined by the ratio between matter and radiation energy densities and sets the location of the matter-radiation equality in the time coordinate. Therefore, the shape--dependent part of $\xi_{l,n}(r)$ is well determined, in a model independent way,  when both $n_S$ and $\omega_m$ are determined at the last scattering surface. We thus marginalise our constraints over the growth parameters with ($n_S$,$\omega_m$) using WMAP7 priors on $n_S$ and $\omega_m$, $n_S=0.963\pm 0.014$ and $\omega_m=0.1334^{+0.0056}_{-0.0055}$.

\subsection{Non--linear suppression cut--off}

On small scales, there are non--linear effects that are not easily modelled via analytical methods, e.g., the internal velocities of galaxies within clusters which cause the famous ``Fingers--of--God" effect seen in galaxy redshift surveys~\cite{1972MNRAS.156P...1J}. We investigate here the effect of such non-linearities and outline our strategy for mitigating them in our analysis.

First, the contribution of velocity--velocity correlation is decomposed from the observed correlation function using the anisotropic feature of redshift distorted galaxy maps of which information is most extractable at the quadrupole moment. Since the leading order of the quadrupole moment is the density--velocity correlation, what we measure is the cross-correlation between the density and velocity fields not the velocity--velocity component. The velocity--velocity correlation is the leading order at the hexadecapole moment, which is too weak to provide tight constraints on the coherent velocity flows. However, both fields (density and velocity) are perfectly correlated in the linear regime at least at scales $k<0.1\hompc$. 

Deviations from a perfect cross--correlation coefficient $\epsilon=1$ ($\epsilon\equiv P_{g\Theta}/\sqrt{P_{gg}P_{\Theta\Theta}}$) are not caused by cosmological effects but rather by non--linear physics. If we analyse our data only in the linear regime (where $\epsilon=1$), then $g_b$ and $g_{\Theta}$ can be simultaneously determined from just the density--density and density--velocity components. Therefore, we impose a cut--off scale on $r=\sqrt{\sigma^2+\pi^2}$ at the scale where $\epsilon$ does not significantly depart from unity. This corresponds to $k<0.1\hompc$, or $r_{\rm cut}=15\mpcoh$, as seen in N--body simulations~\cite{White:2008jy}. We use $k\sim 1/r$, instead of $k\sim 2\pi/r$, given by the peak contribution of convolution in Eq.~\ref{eq:jl} due to the Bessel function in the integrand.

Secondly, we also discard more bins along the $\pi$ direction as they are more contaminated by the ``Finger--of--God" effect. In our model, we use a linear power spectrum instead of a non-linear power spectrum as in~\cite{Scoccimarro:2004tg,Taruya:2009ir,Jennings:2010uv}. This should be fine for the accuracy required in this paper as the most important correction is due to the dispersion effect, which we have included here, and also the effect of non-linearities in the power spectra is only a few percent at $k<0.2 \hompc$ ~\cite{Taruya:2009ir,2010arXiv1006.0699T}. To estimate the impact of the non--linearity on the two--point correction function in redshift space, we use the HALOFIT model~\cite{Smith:2002dz} for the non-linear power spectrum with a CDM transfer function~\cite{Eisenstein:1997jh}. We apply the same non--linear power spectrum for the velocity power spectrum, although this obviously overestimates the velocity power (see ~\cite{Jennings:2010uv} for a non--linear fitting function of the velocity power spectrum). However, we use this non--linear model just for discarding small--scale data which may be significantly affected by non--linearities, and our rejection will be conservative. We find that the effect of non-linearity is over 20$\%$ at the scale of $\sigma < 5\mpcoh$ in the redshift range of our data, and discard these measurements from our analysis. Shown in the right panel of Fig.~\ref{fig:sigma_pi}, measured $\xi_s(\sigma,\pi)$ at bins of $\sigma>5\mpcoh$ is consistent with theoretical predictions (compared with the left panel of Fig.~\ref{fig:sigma_pi}), but $\xi_s(\sigma,\pi)$ at bins of $\sigma<5\mpcoh$ does not match well with the theoretical predictions from the modified Kaiser effect. In this paper, we thus impose a cut--off of $\sigma>5\mpcoh$.

In summary, we impose two cut-offs in scale to control the contamination of small-scale non-linearities on our larger scale correlation function measurements. We achieve this by not including bins with scales $r_{\rm cut}<15\mpcoh$ and  $\sigma<5\mpcoh$ in our fitting procedure.

\begin{center}
\begin{table*}
\begin{tabular}{|c|c|cc|c|cc|}
\hline
 & \multicolumn{3}{|c|}{$z=0.25$} &\multicolumn{3}{|c|}{$z=0.38$}\\
\hline
 & ($\Lambda$CDM) & Kaiser limit  &Modified Kaiser &  ($\Lambda$CDM) & Kaiser limit  &Modified Kaiser \\
\hline
{\bf $g_{\Theta}$}  &  (0.411) &  $0.380^{+0.057}_{-0.056}$  &  $0.377^{+0.056}_{-0.057}$ & (0.422)&
$0.478^{+0.061}_{-0.060}$  &  $0.462^{+0.059}_{-0.059}$  \\

{\bf $\sigma_v$} in Mpc/h  &  (3.28) &$3.04^{+0.46}_{-0.45}$  &  $3.01^{+0.45}_{-0.46}$ & (3.38)&
  $3.82^{+0.49}_{-0.48}$  &  $3.69^{+0.47}_{-0.47}$  \\

{\bf $aH\sigma_v$}  in km/s &  (294) & $272^{+41}_{-40}$  &  $270^{+40}_{-41}$   & (293) &
 $331^{+42}_{-42}$  &  $320^{+41}_{-41}$   \\

\hline
{\bf $g_b$ }  &   & $1.33^{+0.044}_{-0.043}$  &  $1.25^{+0.060}_{-0.062}$  &  &
  $1.32^{+0.047}_{-0.046}$  &  $1.18^{+0.072}_{-0.081}$  \\

{\bf $\beta$} &  &$0.29^{+0.044}_{-0.043}$ & $0.30^{+0.047}_{-0.048}$&  &
$0.36^{+0.048}_{-0.048}$ & $0.39^{+0.056}_{-0.056}$\\

$b_{\rm \Lambda CDM}$   &  & $1.98^{+0.066}_{-0.065}$  &  $1.86^{+0.089}_{-0.093}$  &  &
  $2.09^{+0.074}_{-0.073}$  &  $1.88^{+0.11}_{-0.13}$  \\

\hline
\end{tabular}
\caption{The measured values  of several common cosmological parameters, including $g_{\Theta}$, $g_b$ and $\sigma_v$ which are discussed extensively throughout this paper. The quoted errors on the best fit parameters are one sigma, after marginalizing over all other parameters.  Parameter values given in parenthesis are theoretical predictions based on a  WMAP7--normalized $\Lambda$CDM model for the Universe.}
\label{tab:vp}
\end{table*}
\end{center}

\section{Measured coherent motions}

\subsection{Consequences of the correction to the Kaiser limit}

Using the data and method outlined in Sections II and III, we measure $g_b$  and $g_{\Theta}$ simultaneously from the SDSS DR7 LRG correlation functions assuming the appropriate WMAP7 priors as discussed above. As illustrated in Fig.~\ref{fig:sigma_pi}, there is excellent visual consistency on large scales between our predicted $\xi_s(\sigma,\pi)$ and the observed functions. In detail, we obtain a reduced $\chi^2$ of 0.89 and 0.83 for the $z=0.25$ and 0.38 samples for 12 and 9 degrees of freedom. We have 16 and 13 eigenmodes after truncation, minus four fitting parameter respectively. In Fig.~\ref{fig:contour}, we present constraints on $g_b$  and $\sigma_v$ ($\sigma_v$ is converted from measured $g_{\Theta}$ using Eq.~\ref{eq:sigma_v}). In the Kaiser limit, variations of the coherent growth function of galaxy density fields amplifies mainly the monopole moment of $\xi_s(\sigma,\pi)$, and variations of the coherent motions affects mostly the anisotropy of $\xi_s(\sigma,\pi)$. Measured $g_b$  and $g_{\Theta}$ using the Kaiser limit are presented as unfilled black contours of Fig.~\ref{fig:contour}.

The velocity dispersion in the correlation function $\xi_s(\sigma,\pi)$ of the modified Kaiser effect (Eq.~\ref{eq:modxis}) induces an additional suppression to the $\xi_s(\sigma,\pi)$ of the conventional Kaiser limit (Eq.~\ref{eq:xis}). The variation of coherent motion leads to monopole suppression as well as anisotropic amplification. As coherent motions increase, the corresponding $g_b$ becomes smaller due to the increasing suppression by the velocity dispersion effect. Shown as filled blue contours of Fig.~\ref{fig:contour}, the best fit value of $g_b$ is shifted to lower values and the contours are elongated to smaller $g_b$ at higher $\sigma_v$. While the measured difference of $\sigma_v$ between the Kaiser limit and modified Kaiser effect is less than $4\%$, the monopole shift of $g_b$ due to velocity dispersion with the modified Kaiser effect causes measured differences of $g_b$ from $10\%$ to $15\%$ presented in Table ~\ref{tab:vp}.

\subsection{Cosmological constraints}

We present herein a measurement of the coherent motions using $\sigma_v$, converted from our measured value of  $g_{\Theta}$ (shown in Table~\ref{tab:vp}) and shape factor $D_m$ given by our WMAP7 priors. At $z=0.25$, $\sigma_v$ is measured to be $3.01^{+0.45}_{-0.46}\mpcoh$ (one sigma errors), and at $z=0.38$, we find $\sigma_v=3.69^{+0.47}_{-0.47}\mpcoh$. These measurements are less dependent on the cosmological model for the late--time acceleration of the expansion of the Universe (e.g. dark energy) as our CMB priors are fixed at a much earlier epoch in the Universe, e.g. we use value of $A_S^*$ and $\Omega_m^*$ constrained at the surface of last scattering. Our measurements are also independent of galaxy bias which is a major advantage compared to other methods to parameterise the motions of galaxies from redshift--space distortions. It is true that our measured $\sigma_v$ correlates with $g_b$, but we do not need to know how to separate $b$ and $g_{\delta_m}$ for determining the coherent motions. 

If we assume $H(a)$ is well--measured, then we can convert the units of $\sigma_v$  using $aH\sigma_v$ (assuming $\Lambda$CDM) to give $270^{+40}_{-41}$ km/s and $320^{+41}_{-41}$ km/s at the two redshifts listed in Table~\ref{tab:vp}. Although we would not recommend using $aH\sigma_v$ for cosmological constraints (without a reliable $H$ measurement to an accuracy of a few percent), this conversion is more intuitive when discussing coherent motions and allows us to compare our values with other peculiar velocity measurements, e.g., the ``bulk flow" measurements discussed in Section I.

Traditionally, coherent motions are estimated using the $\beta$ parameter~\cite{1994MNRAS.267..785C,2001Natur.410..169P,2007MNRAS.381..573R,2008Natur.451..541G,2010arXiv1004.3548O},
\ba
\tilde{P}_{\rm ob}(k,\mu)=(1+\beta\mu^2)P_{\delta_g\delta_g}(k)\,.
\ea
This parameter is equivalent to $g_{\Theta}/g_b$ in our notation, and measured to be $\beta=0.30^{+0.047}_{-0.048}$ and $0.39^{+0.056}_{-0.056}$ at $z=0.25$ and 0.38 respectively. As we have measured, $g_b$ is significantly sensitive to the different redshift--space distortion models, and $\beta$ is found to be $\beta=0.29^{+0.044}_{-0.043}$ and $0.36^{+0.048}_{-0.048}$ for the Kaiser limit. Thus, when measuring $\beta$ from the SDSS DR7 LRG samples for use as cosmological constraints, it should be stated clearly which model for redshift-space distortions is assumed.

If we assume a $\Lambda$CDM model, then the galaxy bias can be estimated from our measured $g_b$ values. We present values for $b$ in Table~\ref{tab:vp}, which are $b_{\rm \Lambda CDM}=1.86^{+0.089}_{-0.093}$ and $1.88^{+0.11}_{-0.13}$ at $z=0.25$ and 0.38 respectively. It is interesting to compare our measurements with $b_{\rm \Lambda CDM}=1.86\pm0.07$ as measured by ~\cite{2007MNRAS.378.1196K} using the higher--order correlation function of LRGs from the SDSS. This consistency supports our modified Kaiser formulation, as the estimated $b_{\rm \Lambda CDM}$ using the original Kaiser limit is different by more than one sigma.

\subsection{Tracing the history of coherent motions of galaxies}

\begin{figure}[t]
 \begin{center}
 \epsfysize=3.2truein
 \epsfxsize=3.2truein
   \epsffile{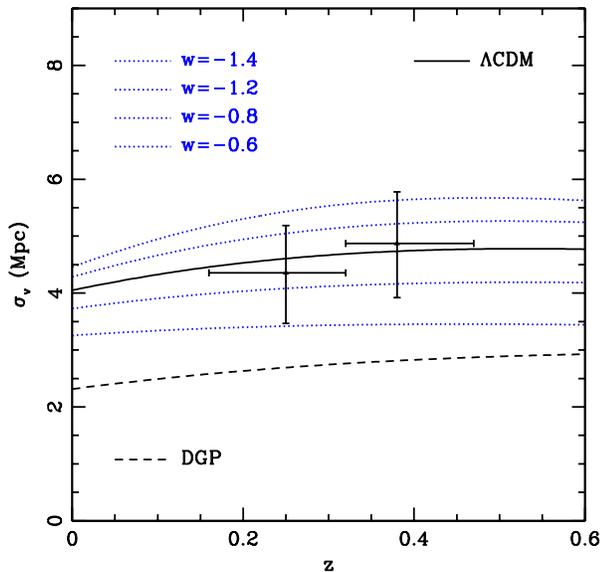}
   \caption{\label{fig:vphistory} \footnotesize We present the measured evolution of coherent galaxy motions at the mean $z=0.25$ and 0.38 which are consistent with predictions for a WMAP7--normalised $\Lambda$CDM model (solid curve). The plotted errors are one sigma as shown in Table I. Dotted curves represent predictions from dark energy models with constant $w=-1.4,-1.2,-0.8$ and $-0.6$ from top to bottom, and dash curve is for DGP model.}
\end{center}
\end{figure}

In Fig.~\ref{fig:vphistory}, we present our measurements of $\sigma_v$, the coherent motions of galaxies in redshift space, versus redshift and compared to theoretical predictions. As can be seen, our measurements are fully consistent with the WMAP7--normalised $\Lambda$CDM model. This represents a ``clean" test of these cosmological models free from contamination by non--linear physics on small--scales and uncertainties of the galaxy bias determination. These data probe the growth history of fluctuations in the Universe and are thus complementary to the geometrical probes of the Universe.


In Fig~\ref{fig:vphistory}, we also present predicted curves for the coherent motion of galaxies for a variety of dark energy models with varying constant equation of state from $w=-1.4$ to $-0.6$ (dotted curves). All other cosmological parameters were fixed to be the same as the WMAP7 best fit $\Lambda$CDM values, except $w$. Using these curves, we can approximately estimate that we have a constraint of $\sigma(w)\sim 0.2$, around a mean value of $w\simeq -1$, from our $\sigma_v$ measurements. This is consistent and complementary to similar constraints from geometrical observations of the Universe~\cite{Lampeitl:2010zx}.

Finally, we also provide in Fig~\ref{fig:vphistory} a prediction for the Dvali-Gabadadze-Porrati (DGP) self-accelerating braneworld scenario~\cite{Dvali:2000hr}. As can be seen, this cosmological model provides a poor description of our observations (excluded at confidence of $3.8\sigma$ for both redshift bins), which support other observational constraints on this model from the cosmic microwave background (CMB) anisotropy, supernovae and Hubble constant data~\cite{Fang:2008kc}.

\section{Discussion and Conclusion}

We present in this paper new measurements of the coherent motions of galaxies on large-scales derived from a new methodology for analysing the redshift--space distortions seen in the observed two--dimensional two--point correlation function of Luminous Red Galaxies from the SDSS DR7 sample. Our new methodology is based on measuring the scale--independent growth functions, $g_b$ (galaxy density) and $g_{\Theta}$ (velocity density), which do not depend upon the physics of the late universe (e.g. dark energy). They do depend on our knowledge of the early Universe, but this can be constrained using information from the Cosmic Microwave Background.

We have determined values of $g_{\Theta}$ from the redshift--space distortions seen in the SDSS DR7 LRG data and, converting these values into the 1--D velocity dispersion $\sigma_v$, we find $\sigma_v=3.01^{+0.45}_{-0.46}\mpcoh$ at a mean redshift of $z=0.25$ and $\sigma_v=3.69^{+0.47}_{-0.47}\mpcoh$ at $z=0.38$. These values for $\sigma_v$ are fully consistent with a WMAP7--normalized $\Lambda$CDM model with $w\simeq-1\pm0.2$ as illustrated in Fig ~\ref{fig:vphistory}. Our observations are however, inconsistent with a DGP model for the Universe to high statistical significance ($>5\sigma$). Our results provide a competitive, and complementary, constraint on these cosmological models compared to the usual geometric probes of the Universe.

We have converted our measured values of $\sigma_v$ into velocity units (as opposed to lengths presented above) and find $270^{+40}_{-41}$ km/s and $320^{+41}_{-41}$ km/s at a mean redshift of $z=0.25$ and $0.38$ respectively, assuming a $\Lambda$CDM Universe. As expected, these coherent motions (or velocity dispersions) are fully consistent with expectations from a $\Lambda$CDM Universe. These estimates are however, inconsistent with local measurements of the peculiar velocity field (or ``bulk flows") which have recently been measured to be greater than these velocities and expectations from $\Lambda$CDM~\cite{Watkins:2008hf,Feldman:2009es}. If the amplitude of these local, observed bulk flows were converted (using $aH\sigma_v$ and assuming a flat $\Lambda$CDM model) to the redshift range studied here ($0.16<z<0.47$), then we might expect to see larger coherent motions.

It is difficult to perform a direct comparison of these different velocity measurements because of the different methods and redshift intervals used. For example, the lower redshift measurements of \cite{Watkins:2008hf,Feldman:2009es} or \cite{Kashlinsky:2008ut,Kashlinsky:2009dw} only probe the velocity field out to 150$h^{-1}$Mpc using their COMPOSITE sample, while most of the data is within a sphere of radius of $\simeq 60h^{-1}$Mpc depending on the weighting scheme used (see Fig 2 of \cite{Feldman:2009es}). In contrast, our statistical estimate of coherent motions on the scales up to 60$h^{-1}$Mpc are derived from within a large volume of the Universe at higher redshift, e.g., 0.5$h^{-3}$Gpc$^3$ ($0.16<z<0.32$) and 1.1$h^{-3}$Gpc$^3$ ($0.32<z<0.47$) respectively, assuming a flat WMAP7 cosmology. Therefore, our measurements have averaged over many hundreds of subregions of approximately the same size as the volume used by ~\cite{Watkins:2008hf,Feldman:2009es} (assuming a subregion of radius of 60$h^{-1}$Mpc). 

Furthermore, there are potential differences in the possible directions of the velocity measurements being compared. The direction of the local velocity measurements of ~\cite{Watkins:2008hf,Feldman:2009es} point towards $\simeq 158, -51$ degrees of Right Ascension (RA) and Declination respectively on the sky, while our SDSS DR7 data is centred at $\sim 180, +30$ (Equatorial coordinates) and concentrated in the northern hemisphere, i.e. our patch is not along the direction relevant to the patches used in~\cite{Watkins:2008hf,Feldman:2009es} or~\cite{Kashlinsky:2008ut,Kashlinsky:2009dw}. A direct comparison of the directions of the various velocity measurements is hard as again our SDSS measurement is a statistical average over many subregions of space and thus has no directional information. That said, our statistical measurement of coherent velocities within such sized volumes of the universe is smaller, to high statistical significance, than that measured locally around our Galaxy.

A more interesting comparison would be with the measurements of \cite{Kashlinsky:2008ut,Kashlinsky:2009dw} who find a ``bulk flow" of X--ray clusters,  with respect to the CMB, of $\simeq1000$ km/s out to $\simeq800h_{70}^{-1}$Mpc (or 560$h_{100}^{-1}$Mpc for comparison herein) in approximately the same direction as discussed above for the local measurements ~\cite{Watkins:2008hf,Feldman:2009es}. Again, it is hard to make a direction comparison in terms of the bulk velocity amplitude and direction as our statistical measurements are derived from higher redshift~\cite{Kashlinsky:2009dw}, they note that their bulk flow velocities peak at $z\le0.16$ with the possibility of higher redshift clusters providing little to their dipole measurements. This is therefore, below our low redshift LRG sample data, and averaged over a larger volume, than used by \cite{Kashlinsky:2008ut,Kashlinsky:2009dw}.

One possible explanation for the differences we are seeing is that our Galaxy is located in an unusual part of the Universe, e.g., in a highly over, or underdense region of the Universe. Again, our measurements are obtained at $z>0.16$, beyond the redshift limits of all these local measurements~\cite{Watkins:2008hf,Feldman:2009es,Kashlinsky:2008ut,Kashlinsky:2009dw}. Moreover, we can not exclude the presence of a large constant large-scale ``dark flow" across the volume surveyed by our DR7 data i.e., a velocity dipole with no variation across $\simeq1h^{-3}$Gpc$^3$ of the Universe. Such a constant flow would be undetectable by our method as the correlation function in redshift space is distorted by the divergence of the peculiar velocity field, while measuring the bulk flows via the SZ effect can include a possible global flow. However, this explanation would require the ``dark flow" of \cite{Kashlinsky:2009dw} to extend out beyond the SDSS LRG sample ($z\simeq0.5$) to leave it undetectable, i.e., over a 1000$h^{-1}$ Mpcs in scale. This can be tested using the SDSS-III Baryon Oscillation Spectroscopic Survey (BOSS), which will provide redshift-space distortion measurements both at higher redshift and over larger volumes of the Universe~\cite{White:2010ed}.

In the future we plan to extend our measurements to higher redshifts, in order to extend the history of coherent motions (e.g. Fig. \ref{fig:vphistory}). As shown in our paper, these motions can be measured in an independent way, free of some of the problems associated with other measures of the growth history of the Universe, and the assumed cosmological model for the late--time Universe, e.g., we are able to test not only the conventional dark energy model, but also ``interacting" and ``clustered" dark energy models, not to mention the general class of modified gravity theories. Unlike other approaches, our measurements are free from any possible violation of the consistent equations.

\section*{Acknowledgments}

We thank the referee for helpful comments that improved the quality of this paper. Y-SS, CGS and RN are supported through the STFC, and I.K. acknowledges support by JSPS Research Fellowship and WPI Initiative, MEXT, Japan. We thank Eyal Kazin for significant assistance in using his LRG data sample, as well as many fruitful discussions, and Atsushi Taruya for helpful comments in formulating Eq.~17. We also thank Ofer Lahav for fruitful discussions. The authors acknowledge the use of the UCL Legion High Performance Computing Facility, and associated support services, in the completion of this work. We thank Korea Institute for Advanced Study for providing computing resources (KIAS linux cluster system) for this work.


\end{document}